# Out-of-plane transport of 1T-TaS$_2$/graphene-based van der Waals heterostructures

Carla Boix-Constant,[1] Samuel Mañas-Valero,[1,*] Rosa Córdoba,[1]
José J. Baldoví,[1] Ángel Rubio,[2,3] Eugenio Coronado[1,*]

[1] *Instituto de Ciencia Molecular (ICMol), Universitat de València, Catedrático José Beltrán Martínez n° 2, Paterna 46980, Spain.*
[2] *Max Planck Institute for the Structure and Dynamics of Matter and Center for Free-Electron Laser Science, Luruper Chaussee 149, 22761, Hamburg, Germany.*
[3] *Nano-Bio Spectroscopy Group, Departamento de Física de Materiales, Universidad del País Vasco, 20018 San Sebastian, Spain.*

**E-mail:** samuel.manas@uv.es, eugenio.coronado@uv.es

**Keywords:** 2D materials, van der Waals heterostructures, Quantum materials, electrical properties, DFT calculations

Due to their anisotropy, layered materials are excellent candidates for studying the interplay between the in-plane and out-of-plane entanglement in strongly correlated systems. A relevant example is provided by 1T-TaS$_2$, which exhibits a multifaceted electronic and magnetic scenario due to the existence of several charge density wave (CDW) configurations. It includes quantum hidden phases, superconductivity and exotic quantum spin liquid (QSL) states, which are highly dependent on the out-of-plane stacking of the CDW. In this system, the interlayer stacking of the CDW is crucial for the interpretation of the underlying electronic and magnetic phase diagram. Here, thin-layers of 1T-TaS$_2$ are integrated in vertical van der Waals heterostructures based on few-layer graphene (FLG) contacts and their electrical transport properties are measured. Different activation energies in the conductance and a gap at the Fermi level are clearly observed. Our experimental findings are supported by fully self-consistent DFT+U calculations, which evidence the presence of an energy gap in the few-layer limit, not necessarily coming from the formation of out-of-plane spin-paired bilayers at low temperatures, as previously proposed for the bulk. These results highlight dimensionality as a key effect for understanding the properties of 1T-TaS$_2$ and opens the door to the possible experimental realization of low-dimensional QSLs.

## 1. Introduction

Low-dimensional materials offer unique possibilities for studying strongly correlated materials with tantalizing physical phenomena such as superconductivity or magnetism.[1–3] A relevant example in this context is provided by transition metal dichalcogenides (TMDCs). These compounds, with general formula $MX_2$ (where M is a transition metal and X is a chalcogen), are formed by the stacking of X-M-X layers united through van der Waals interactions (see **Figure 1.a**).[4–7] The physical properties of these layered materials range from insulators to superconductors and, as recently discovered, some of them show intrinsic magnetic properties.[8,9]

In particular, 1T-TaS$_2$ is one of the most studied TMDCs due to its unexpected physical properties. Like the rest of group V TMDCs, this compound should be in theory a metal. However, it behaves as a semiconductor and does not show any magnetic signature of long range order even at milliKelvin temperature.[10,11] Indeed, thinking about these non-conventional properties, Philip W. Anderson conceived his resonance band model.[12] One key element for the richness of its electronic properties is due to the formation of charge density waves (CDW) –periodical localizations of the charge– that are stable with different arrangements, yielding to metastable phases, quantum hidden states[13] and even quantum spin liquid (QSL) phases [11,14,15] and superconductivity when doped or under pressure.[16–21] Upon cooling, the thermal behavior of bulk 1T-TaS$_2$ shows the formation of an incommensurate charge density wave (I-CDW) at 350 K, followed by a nearly commensurate CDW (N-CDW) to a commensurate CDW (C-CDW) transition around 200 K, generally ascribed as a Mott transition.[22] The previous transitions are accompanied by a characteristic hysteresis behavior, as shown by transport measurements or magnetic susceptibility.[23] The CDW arranges forming the so-called Star-of-David (SD), where every 13 Ta atoms are coupled: 12 of them pair (marked in brown in **Figure 1.b** and **Figure 1.c**) and shift towards the central one (marked in red in **Figure 1.b** and **Figure 1.c**), forming a triangular lattice of S = ½ electrons below 200 K (**Figure 1.b**).[24] This triangular lattice is magnetically frustrated due to the existing antiferromagnetic correlations and is the basis for the possible emerging QSL, as proposed both theoretically[14] and experimentally by different techniques.[11,15,25]

In the monolayer limit, the QSL ground state is theoretically demonstrated[14] but, in the multilayer case, the out-of-plane pairing of the SD has to be considered (see **Figure 1.c**). For instance, if out-of-plane correlations between the layers are absent (A or L stacking, **Figure 1.c**), then the Mott scenario is valid and the QSL is formed; meanwhile, if the SDs form dimers in the out-of-plane direction as paired bilayers (AL stacking, **Figure 1.c**), the system has just to be considered as a conventional band insulator –thus, not being a QSL–.[14,26] Nonetheless, these options do not need to exclude each other necessarily since they can occur at different temperature regimes.[27] Thus, studying the out-of-plane correlations is a key element for understanding the underlying electronic and magnetic scenario occurring in 1T-TaS$_2$.

Notice that so far most of the studies on 1T-TaS$_2$ have been focused on bulk samples, both from the theoretical and experimental points of view and the out-of-plane metallic character of the bulk is still under debate. From the experimental side, different interlayer mechanisms have been proposed and the conventional Mott picture, dimerization and the formation of domain wall networks, among others, are being revisited.[27–31] Theoretically, some authors have proposed metallic band dispersion along the out-of-plane direction if the SD are coupled vertically (A stacking, **Figure 1.c**), or through a diagonal (L stacking, **Figure 1.c**) and an insulating gap in the case of the formation of bilayers (AL stacking, **Figure 1.c**).[32,33] However, more recent calculations applying a self-consistent DFT+U generalized basis approach that covers the whole SD point towards a Mott gap in the bulk, independently of the out-of-plane stacking configuration of the SDs.[34]

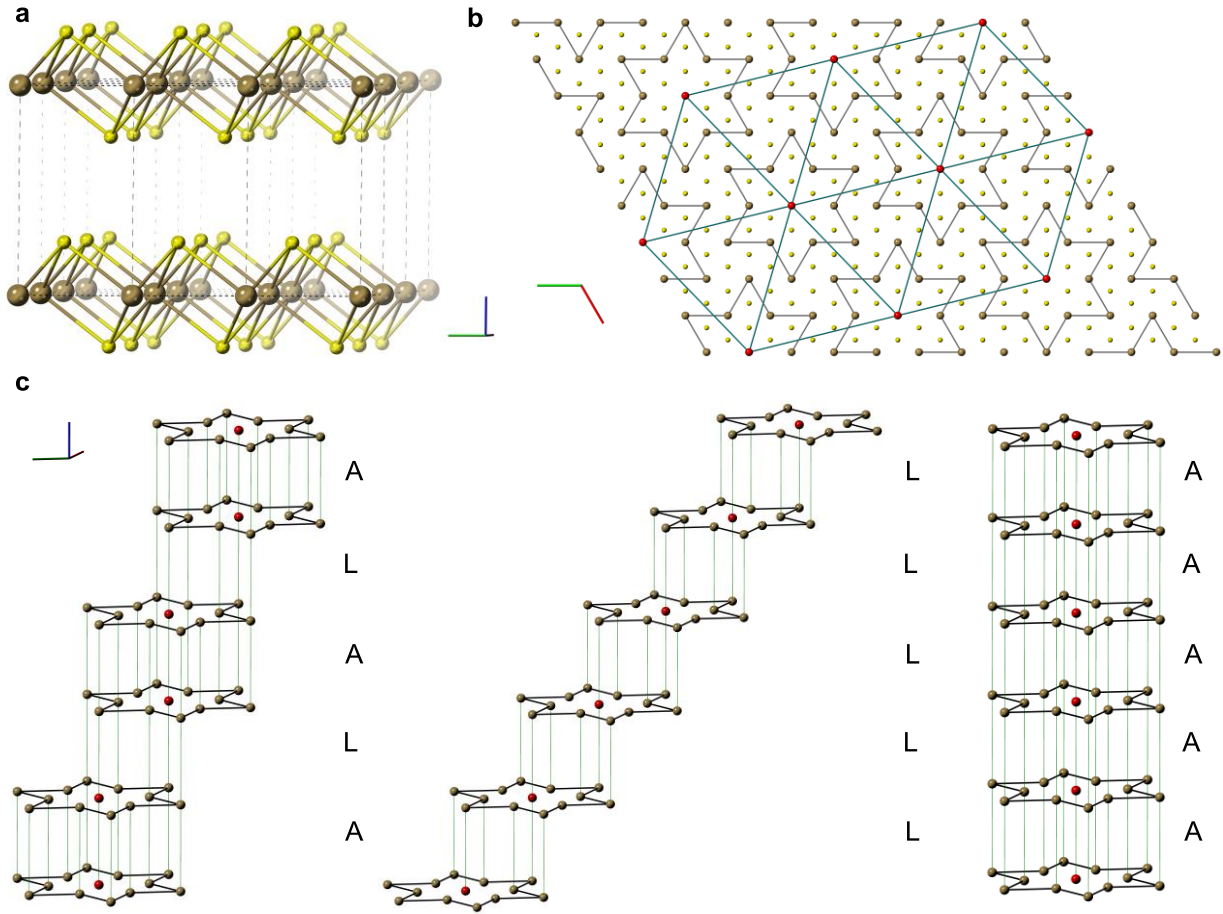

**Figure 1.** Crystal structure and CDW arrangement in 1T-TaS$_2$. (a) Unit cell in dash lines. (b) In-plane structure where it has been highlighted the formation of the stars of David (SD), corresponding to the commensurate charge density wave phase (grey line) as well as the triangular lattice of spin ½ (turquoise line) that forms the quantum spin liquid. (c) Different out-of-plane stacking configurations of the CDW: AL stacking (left), L stacking (center) and A stacking (right), following the notation of reference [33]. Sulphur atoms are represented in yellow and tantalum ones in brown. For clarity, the central tantalum of the SD (S = ½) has been marked in red and the different stacking configurations with green lines. The red, green and blue colors of the axis correspond to the a, b and c axis, respectively. For simplicity, in c it has been only represented the tantalum atoms.

It is worthwhile to mention that the previous discussions are hold regarding bulk 1T-TaS$_2$ but atomically-thin layers do not necessarily behave in the same way as a consequence of their reduced dimensionality. In the thin-layer regime (thickness below 10 nm), several experiments have been performed in order to unveil the CDW phase diagram of 1T-TaS$_2$.[20,22] In fact, the fine tuning of the CDW has been achieved by using conventional electrodes and applying current pulses,[35] bias voltage,[36] light,[37,38] strain,[39] or its deposition on different substrates,[40] among others. With devices based on van der Waals heterostructures (vdWHs), the few examples reported so far have involved transport measurements on thin-layers of 1T-TaS$_2$ contacted to graphene,[41] MoS$_2$[42] or black phosphorus[43] in a typical in-plane configuration. Only bulky samples (thicknesses of hundreds of nanometers) [28,29,44] or devices incorporating superconducting NbSe$_2$[45] have been measured in a vertical configuration. Hence, the out-of-plane CDW structure of 1T-TaS$_2$ in the thin-layer limit remains unexplored. Thus, in this work we present the fabrication and electrical characterization of vertical vdWHs based on 1T-TaS$_2$ thin-layers interfaced with few-layers graphene (FLG).

## 2. Results and discussion

The vdWHs fabrication is based on the successive deterministic transfer of the 2D materials and is implemented inside an argon glovebox.[46] From the mechanical exfoliation of the flakes and its inspection to the vdWH assembly and electrical transport measurements, the vdWHs are not exposed to air (see methods for further details). This is a key factor since thin-layers of 1T-TaS$_2$ degrade in ambient conditions.[20,35] The identification of thin-layers of 1T-TaS$_2$ on 285 nm SiO$_2$/Si substrates is performed by optical microscopy. The optimal contrast region is observed for the green channel of the visible spectrum, as it is experimentally and theoretically investigated within the Fresnel framework in the **Supplementary Section 1**. The vdWH consists on a 1T-TaS$_2$ thin-layer sandwiched via van der Waals forces between few-layers graphene (FLG) that are placed on top of the metallic contacts (**Figure 2.a** and **Figure 2.b**). Although the geometrical factors cannot be fully controlled within the present vdWH approach, it benefits of being fully integrated under inert atmosphere conditions, thus being a unique excellent option for studying air-unstable 2D materials, as recently performed with some highly-unstable 2D magnets as CrI$_3$.[47] The electrical-transport characterization is performed by standard four-probe configuration: a current is injected through the whole vdWH by the outer leads and the voltage drop is measured using the inner ones (**Figure 2.a** and **Figure 2.b**; see methods for technical details). **Figure 2.d** shows a scanning transmission electron microscopy (STEM) image of an area of the cross-sectional view of the vertical vdWH, where 5 layers of 1T-TaS$_2$ can be clearly identified (see methods and **Supplementary Section 5** for other vdWHs).

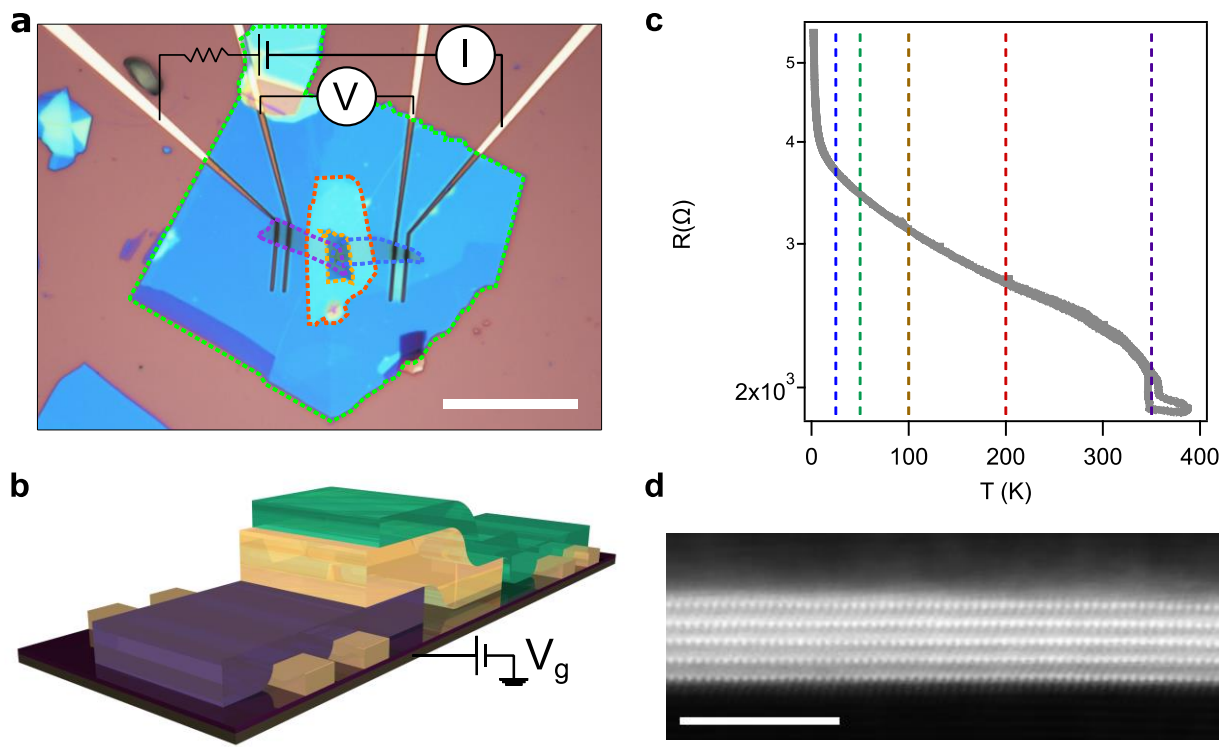

**Figure 2.** Vertical van der Waals heterostructures (vdWH) based on 1T-TaS$_2$. (a) Optical image of the vdWH and electronic transport configuration. For clarity, the different flakes are enclosed with dashed lines: the bottom h-BN in red, the bottom contact in purple, the 1T-TaS$_2$ flake in orange, the top contact in blue and the top h-BN in green. Scale bar: 20 μm. (b) Artistic representation –not to scale– of the vdWH with the back gate voltage configuration sketched. In dark yellow it is represented the metal contacts, in purple the bottom FLG contact, in green the top FLG contact and, in yellow, the 1T-TaS$_2$ thin-layer. (c) Electrical transport properties

of a 9 layers 1T-TaS$_2$ vdWH (device A in the **Supplementary Information**). The dashed lines highlight the different transitions described in the literature for bulk 1T-TaS$_2$ (see main text). In particular, from I-CDW to N-CDW at 350 K (purple), from N-CDW to C-CDW at 200 K (red), from C-CDW to H-CDW at ca. 50 K (green) and the QSL crossovers described at 100 K (yellow) and 25 K (blue). (d) STEM image of a vdWH showing 5 layers of 1T-TaS$_2$ (device D in the **Supplementary Information**). Scale bar: 5 nm.

A total of 9 different vdWHs with 1T-TaS$_2$ thickness ranging from 3 to 9 layers are fabricated and their electrical transport properties inspected (the characterization of all the vdWHs is discussed in the **Supplementary Information**). The electrical properties of a representative device (device A in the **Supplementary Information**) are shown in **Figure 2.c**. As reference, the different transitions reported in the literature for bulk 1T-TaS$_2$ are also plotted in the figures as vertical dashed lines: 350 K (I-CDW),[48] 200 K (C-CDW),[48] 100 K (QSL crossover),[49] 50 K (H-CDW and QSL crossover)[13,15,28,50] and 25 K (QSL crossover).[49] The characteristic hysteretic transition from I-CDW to N-CDW transition at 350 K (**Figure 2.c** and **Supplementary Section 2**) is clearly observed even in our thinnest vdWH (3 layers; **Supplementary Section 2, 4** and **5**), in contrast to previous reports where it was absent for thicknesses below 3.5 nm (*ca.* 6 layers).[20] We attribute it to the fact of working under strict inert atmosphere conditions that avoid the oxidation of 1T-TaS$_2$. While cooling down, the vdWH resistance increases without exhibiting the abrupt hysteresis around 200 K observed in bulk samples (**Figure 2.c** and **Figure S.7**), as attributed to the reduced dimensionality of the 1T-TaS$_2$ thin-layers.[51] In order to relate different devices, the resistance at different temperatures, R(T), is normalized, R$_{norm}$, to the value of the resistance at the highest temperature measured (ca. 400 K), R(HT) (Supplementary Section 3), thus R$_{norm}$ = R(T)/R(HT). In **Figure S.7** and **Figure S.8** (**Supplementary Section 2**), measurements of different horizontal devices are compared to those of vertical ones (this work). In these vertical heterostructures, $R_{norm}$ is *ca.* two times lower in the high temperature regime and several orders of magnitude at low temperatures (2 K), reflecting that the insulating state due to the in-plane commensuration of the SD formation (hysteretic behavior at 350 K) is affecting more the overall transport properties of the horizontal devices. This is in line with transport measurements in bulk 1T-TaS$_2$, which show a much smaller resistance change at 350 K in the out-of-plane direction compared to the in-plane one.[29] From all these observations, it is inferred that the out-of-plane contribution is the dominant term in our vertical devices. All measured vdWHs exhibit similar trends (see **Supplementary Information**).

The vdWH resistance in the present case has two main contributions: the intrinsic 1T-TaS$_2$ resistance (formed by the in-plane and the out-of-plane resistances in parallel) and the contact resistance at the interface, $R_{interface}$ (formed between the thin-layers of FLG and 1T-TaS$_2$).[28] $R_{interface}$ is not expected to exhibit a temperature dependence unless an electronic gap or barrier –as, for example, a Schottky barrier– is formed.[52] A simple way to consider these transitions is by assuming that the transport properties for distinct CDW configurations are associated with different activation energies. This can be modelled by an Arrhenius law, where the conductance, G, is of the form $G = G_0 \cdot \exp(-E_a/k_B T)$, being $G_0$ a pre-factor, $E_a$ the activation energy, $k_B$ the Boltzmann constant and T the temperature.[53] Other mechanisms, as variable range or nearest-neighbor hopping conduction,[54] are also considered (see **Supplementary Section 3**), not observing any single transport mechanism suitable for all the temperature range. In the Arrhenius plots (**Supplementary Section 3**), two linear regimes are investigated: one at low temperatures (LT) and one in the N-CDW region (200 K – 350 K). The present data shows a progressive transition between the HT and LT regimes where different activation energies regimes can be found, without any sharp or abrupt transition below 350 K, in stark contrast with bulk 1T-TaS$_2$.[28] The involved activation energies are in the order of 0.05 meV and 11 meV, for the LT and HT regimes, respectively. The value for the HT phase is comparable with the one reported by Svetin *et al.* [28] for bulky samples (10 meV) in the 40 K – 140 K range. Note that, in contrast to the results reported for bulk 1T-TaS$_2$ in the out-of-plane direction,[28,29] we do not find a linear regime below 200 K, nor the CDW hysteresis at 200 K. This underlines the possible electronic changes due to dimensionality effects (bulk crystals of ~100 μm vs. atomically-thin layers below 10 nm) and differences in the geometrical factors of the devices. The projection of the LT and HT fittings intercepts around 70 K. This temperature is in accordance with that proposed for the H-CDW to C-CDW transition reported by previous photo-transport experiments.[13,28]

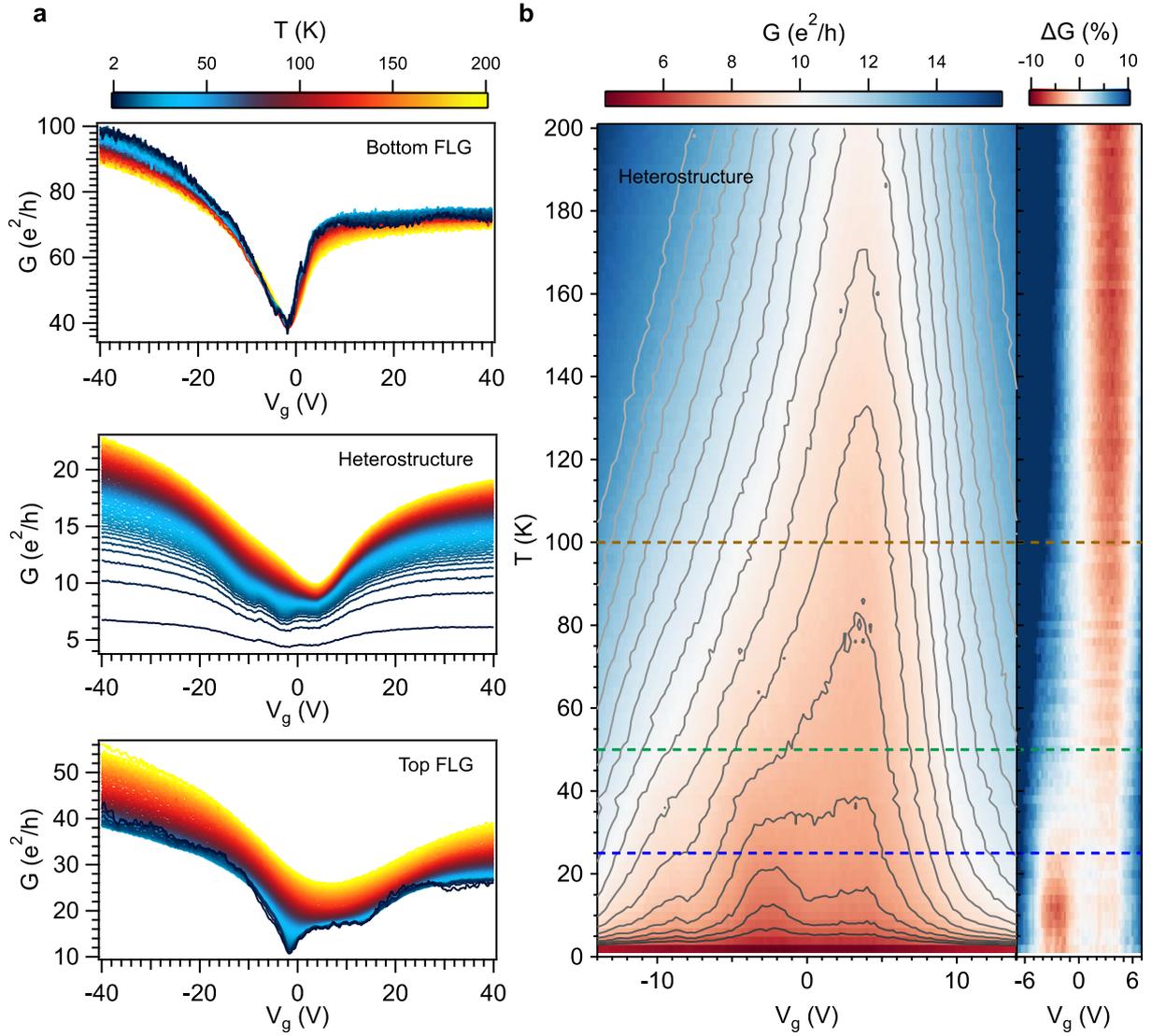

**Figure 3.** Conductance as a function of the back gate voltage of a 9 layers 1T-TaS$_2$ heterostructure (device A in the Supplementary Information). (a) Temperature dependence for the bottom FLG, the heterostructure and the top FLG. (b) Detailed view around $V_g = 0$ for the heterostructure where, for clarity, it has been added a contour plot, the values of $\Delta G = \{[G(V_g) - G(V_g = 0)]/G(V_g) = 0)\} \cdot 100$ and dashed lines for the previous reported transitions in bulk 1T-TaS$_2$ (C-CDW to H-CDW transition at ca. 50 K –green– and the QSL crossovers at 100 K –yellow– and 25 K –blue–).

The use of a back gate voltage, $V_g$, allows the experimental access to the density of states of the vdWH. Although the 1T-TaS$_2$ carrier density is high, requiring in principle the use of electrochemical techniques,[20] it has been shown recently that a solid gate is effective when thin-layers of 1T-TaS$_2$ are integrated in a heterostructure with other 2D materials, like black phosphorous or 2H-MoS$_2$.[42,43,55] As shown in **Figure 3** for the vdWH of 9 1T-TaS$_2$ layers (device A in the **Supplementary Information**), at 200 K the conductance as a function of the back gate voltage presents a minimum that broadens and shifts towards 0 V upon cooling down, until a plateau is reached around 50 K and a second peak emerges at T < 25 K. This second minimum at $V_g \sim -2$ V is related to the charge neutrality point (CNP) of the FLG contacts. The behavior of the overall vdWH cannot be ascribed to the simple addition of the CNP of the bottom and top FLG (**Figure 3.a**), as corroborated by performing DC IV curves at different back gate voltages and temperatures and comparing their resistances (**Supplementary Section 4.1**). For clarity, in **Figure 3.b** the conductance is normalized, $\Delta G$, with respect to the value at zero voltage gate, being $\Delta G = \{[G(V_g) - G(V_g = 0)]/ G(V_g = 0)\} \cdot 100$; thus, $\Delta G < 0$ (red color in Figure 3.b) represents a higher resistance that can be attributed to a lower density of states with respect to the case of $V_g = 0$. Assuming that the Fermi level of the whole vdWH resides at $V_g = 0$, it can be seen that at low temperatures an electronic gap emerges around $V_g = 0$. Since the CNP remains at the same position in the vdWH and in the FLG contacts, the origin of the plateau in conductance cannot be ascribed to different doping levels in the FLG flakes. This gap state shows a magnetic field dependence, as well (**Figure 4**). At 2 K, the gap at $V_g = 0$ survives until a field of *ca.* 1 T and, while increasing the temperature, this value decreases (more detailed data can be seen in the **Supplementary Section 4.1**). At high magnetic fields, the characteristic Landau levels of graphene that develop forming a Landau fan are observed.[56] The overall same tendencies have been observed in several vdWHs (**Supplementary Sections 4.1, 4.2, 4.3 and 4.9**), although the insulating gap size differs from sample to sample. This may be attributed to the different geometrical factors between vdWHs.

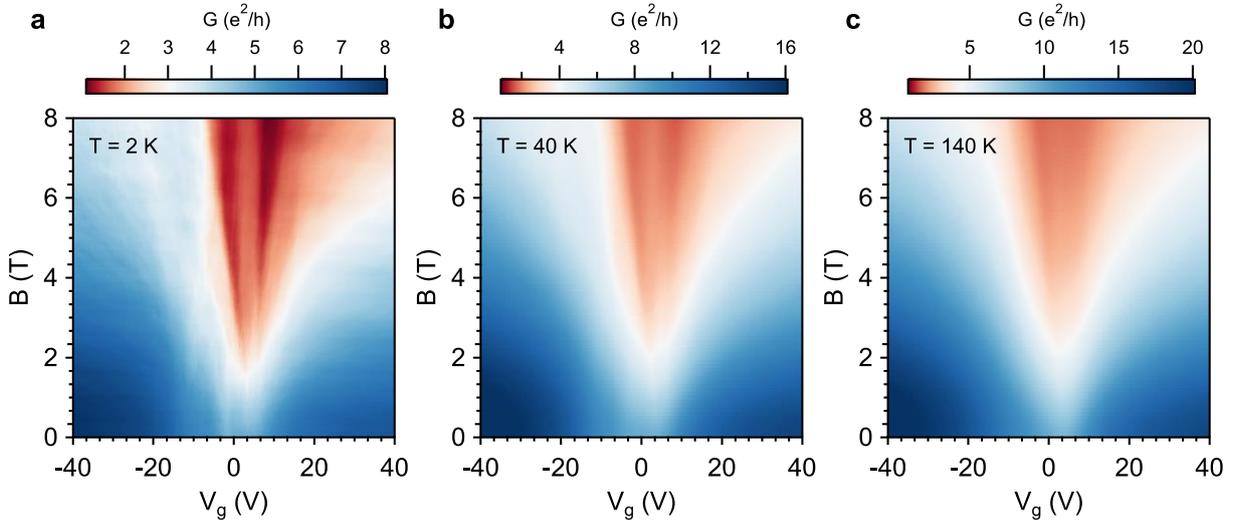

**Figure 4.-** Conductance as a function of the back gate voltage and perpendicular external magnetic field for the 9 layers 1T-TaS$_2$ heterostructure (device A in the Supplementary Information) at 2 K, 40 K and 140 K.

In order to explore the nature of the insulating behavior observed in the conductance experiments, we carry out first principles calculations for different number of 1T-TaS$_2$ layers. Owing to the strong correlation of electrons in the *d* orbitals of the Ta atoms due to the CDW formation, we adopt a spin-polarized DFT+U approach, where *U* is the on-site Coulomb repulsion. We estimate self-consistently a Hubbard *U* of 2.86 eV for the *d* orbitals of the central Ta in the SD using density functional perturbation theory (DFPT)[57] in QuantumEspresso.[58]

This value is in accordance with the obtained U from other authors (2.94 eV) using the same DFPT method for the undeformed structure.[29] The $\sqrt{13}$ x $\sqrt{13}$ supercells for 1 layer, 2 layers, 3 layers, 4 layers and bulk (containing 2 layers) are fully optimized without constrains considering the Hubbard *U* in all cases. Then, we compute the electronic band structure for the different slabs and the bulk, applying *U* to the *d* orbitals of the Ta atoms of the whole SD (**Figure 5**).

In **Figure 5.a** one observes that the in-plane bands (Γ-M-K-Γ) are gapped around the Fermi energy, independently of the number of layers. This is due to the formation of a commensurate CDW in the material, which avoid the in-plane hopping between neighboring SDs. On the contrary, the out-of-plane bands (Γ-A) exhibit marked band dispersion in the bulk in agreement with previous calculations in the literature (**Figure 5.b**),[32–34] displaying a Mott gap (~0.19 eV) that is dependent on *U* and can be suppressed by the application of external stimuli such as moderate hydrostatic pressure.[29] This band dispersion flattens when the dimensionality is reduced and a more robust bandgap appears, which is independent of the SDs stacking and the degree of correlation (see **Supplementary Section 6**), being enhanced when the number of layers is reduced (from 0.19 eV in the 4 layers slab to 0.39 eV in the monolayer, **Figure 5.a**). Therefore, this gap, which is intrinsic to the monolayer, can also be taken as a fingerprint of the few-layers 1T-TaS$_2$ system. It arises from the confinement of the electrons within the SD, either in-plane as well as out-of-plane, as illustrated by the flat character of the band dispersion near the Fermi level in the few-layer limit. This is in sharp contrast with the bulk where a dispersive band near the Fermi level is predicted, thus facilitating the out-of-plane delocalization. This manifests that dimensionality effects play a key role in the electronic behavior of 1T-TaS$_2$ in such a way that the observed insulating behavior is not necessarily due to the formation of out-of-plane spin-paired bilayers at low temperatures, as theoretically proposed for the bulk case.[32,33]

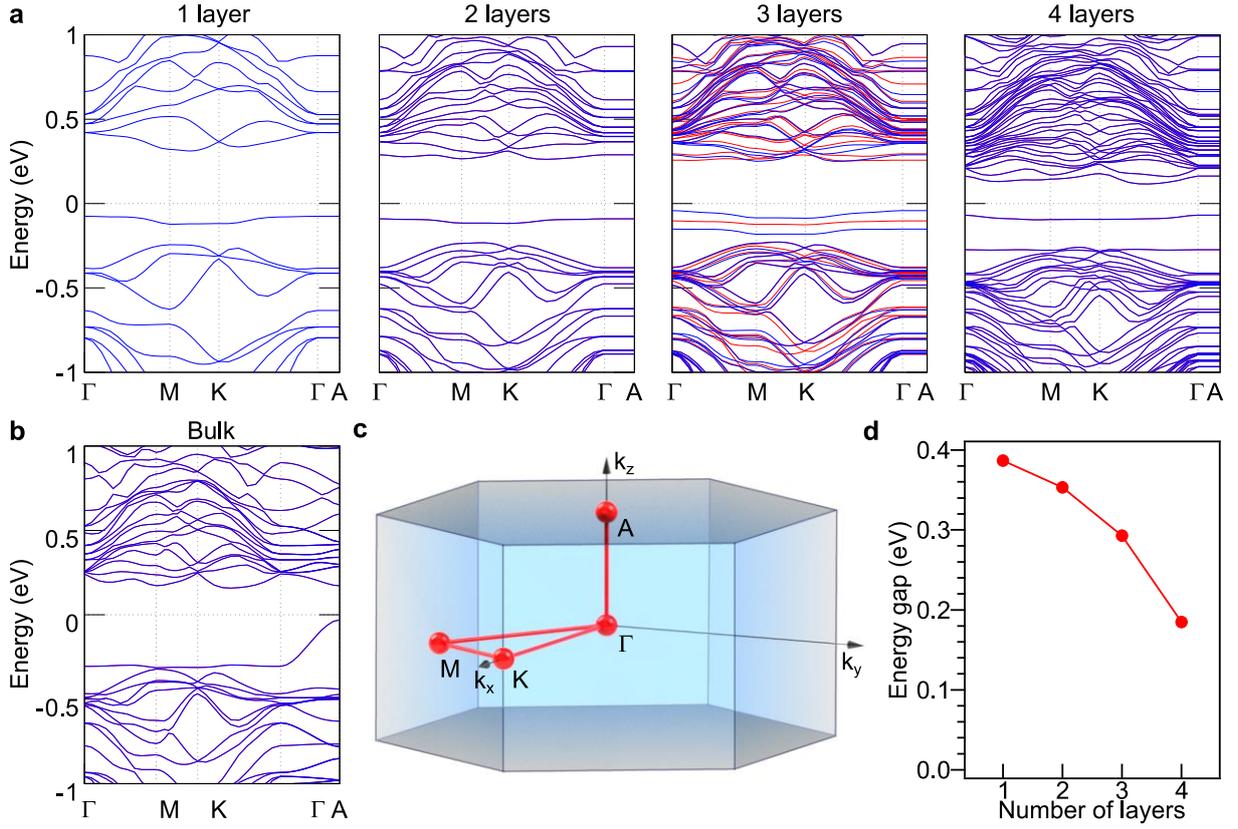

**Figure 5.-** Calculated band structure for C-CDW $\sqrt{13} \times \sqrt{13}$ supercells of 1T-TaS$_2$ for (a) atomically-thin layers (from 1 layer to 4 layers from left to right) and (b) bulk (blue/red refers to spin up/down). The Brillouin zone is depicted in c. (d) Thickness dependence of the gap.

## 3. Conclusion

Different CDW multi-stable configurations are explored in air-unstable thin-layers of 1T-TaS$_2$ by the fabrication of vertical vdWHs with FLG. Based on an Arrhenius model, a progressive transition with different activation energies is observed from 200 K (C-CDW) to the lowest temperature with a slope change at *ca.* 70 K, in agreement with the proposed H-CDW. In addition, a gap in the Fermi level emerges at low temperatures. The present results, supported by fully self-consistent DFT+U calculations, highlight the electron confinement across the layers when approaching to the few layer limit.

So far, the requisites for stabilizing a QSL state in these layered materials are: i) to have a frustrated spin network with antiferromagnetic correlations (encountered here by a commensurate SD structure), ii) to have electron confinement (in-plane as well as out-of-plane localization) and iii) to have an odd number of electrons per unit cell (even numbers, for example bilayers, yield to a conventional band insulator). In the present case the first requirement is fulfilled and we have demonstrated the second one, both experimentally and theoretically. As far as the third condition is concerned, this remains elusive with the techniques used in this work (transport measurements). However, it is worth noting that we have observed the formation of a gap in atomically-thin layers of 1T-TaS$_2$ having an odd number of layers (3, 5 and 9 layers, as proved by cross-section STEM). Hence, even in presence of a bilayer paring, in these specific cases an unpaired layer would remain, where a QSL could exist. This would be highly relevant to the fields of quantum information and communication as it opens the possibility of using these low dimensional 1T-TaS$_2$ materials for fabricating devices based on 2D van der Waals heterostructures displaying large entanglement effects.

## 4. Methods

*Crystal growth of 1T-TaS$_2$:* High quality crystals were grown by Chemical Vapor Transport (CVT) using iodine as a transport agent, as already reported by some of us.[49]

*Exfoliation, characterization and manipulation of atomically-thin layers:* 1T-TaS$_2$ crystals and natural graphite were mechanically exfoliated using adhesive tape (80 μm thick adhesive plastic film from Ultron Systems) inside an argon glove box. The exfoliated samples were inspected primarily by optical microscopy (Nikon Eclipse LV-100 optical microscope with a Nikon TU Plan Fluor 100x objective lens of 1 mm working distance and a numerical aperture of 0.9 and 10 nm FWHM visible band pass filters from Thorlabs) as a fast tool for the identification of thin-layers and atomic force microscopy (Nano-Observer AFM from CSI Instruments) inside an argon glove box. Following the recently developed fabrication of vertical vdWHs for studying air unstable 2D magnets, like CrI$_3$ or MnPS$_3$,[47,59] the vertical vdWHs were built in a deterministic way using polycarbonate, as reported in reference [46] and placed on top of pre-lithographed metal contacts fabricated by conventional electron beam lithography techniques (5 nm Ti/ 50 nm Pd) on 285 nm SiO$_2$/Si substrates (from NOVA Electronic Materials, LLC).

*Electrical measurement set-up:* Electrical measurements were performed in a Quantum Design PPMS-9 with a base temperature of 2 K, using conventional DC and AC lock-in techniques with a MFLI Lock-in Amplifier from Zurich Instruments at low frequencies (27.7 Hz). All shown measurements are performed in DC unless it is explicitly said that AC was applied. In order to perform current-bias experiments, an external resistance of 100 MΩ is used, i.e., much larger than the resistance of the sample. Bottom back gate voltage was applied with a Keithley 2450. All temperature sweeps were performed at 1 K/min and field sweeps at 200 Oe/s.

*TEM:* Standard TEM cross-sectional sample preparation by using a FEI Helios Nanolab 650 Dual Beam instrument was carried out on the vertical vdWHs. Scanning transmission electron microscopy (STEM) imaging was carried out with a probe-corrected FEI Titan 60–300 operated at 300 kV and equipped with a high brightness X-FEG and a Cs CETCOR corrector for the condenser system to provide sub-angstrom probe size. STEM images are shown in **Supplementary Section 5**.

*Band structure calculations:* DFT+U electronic structure calculations were performed using the Quantum ESPRESSO package.[58] Exchange-correlation energy is described considering the Perdew–Burke–Ernzerhof (PBE) generalized gradient approximation (GGA) functional. We use standard solid-state pseudopotentials (SSSP) from the efficiency library of Materials Cloud.[60,61] The electronic wave functions are expanded with well-converged kinetic energy cut-offs for the wave functions and charge density of 45 and 360 Ry, respectively. The Brillouin zone was sampled by a fine Γ-centered $8 \times 8 \times 8$ k-point Monkhorst–Pack mesh for the case of the bulk system and $8 \times 8 \times 1$ for the case of slab calculations. Dispersion corrections to account for van der Waals interactions between the 1T-TaS$_2$ layers are considered by applying semi-empirical Grimme-D3 corrections. All the structures are fully optimized using the Broyden-Fletcher-Goldfarb-Shanno (BFGS) algorithm until the forces on each atom are smaller than $1 \cdot 10^{-3}$ Ry/au and the energy difference between two consecutive relaxation steps is less than $1 \cdot 10^{-4}$ Ry. The Hubbard $U$ is determined self-consistently using used the simplified version proposed by Dudarev et al.[62] using density functional perturbation theory.[57]

**Supporting Information**
Supporting Information is available from the corresponding author.

**Acknowledgements**
We acknowledge the financial support from the European Union (ERC AdG Mol-2D 788222 and ERC-2015-AdG-694097), the Spanish MICINN (MAT2017-89993-R co-financed by FEDER and Excellence Unit "María de Maeztu", CEX2019-000919-M), the Generalitat Valenciana (Prometeo program and PO FEDER Program, ref. IDIFEDER/2018/061 and IDIFEDER/2020/063), the Basque government (Grupos Consolidados, IT1249-19) and the Deutsche Forschungsgemeinschaft (DFG) under Germany's Excellence Strategy - Cluster of Excellence Advanced Imaging of Matter (AIM) EXC 2056 - 390715994 and funding by the Deutsche Forschungsgemeinschaft (DFG) under RTG 1995 and GRK 2247. Support by the Max Planck Institute - New York City Center for Non-Equilibrium Quantum Phenomena is acknowledged. R. C. acknowledges the support of a fellowship from "la Caixa" Foundation (ID 100010434). The fellowship code is LCF/BQ/PR19/11700008. JJB thanks support from the Plan Gent of Excellence of the Generalitat Valenciana (CDEIGENT/2019/022). C.B.-C. thanks the Generalitat Valencia for a PhD fellowship.

We thank Ángel López-Muñoz for his constant technical support and helpful discussions. The electron microscopy measurements have been conducted in the "Laboratorio de Microscopías Avanzadas" at "Instituto de Nanociencia de Aragón - Universidad de Zaragoza" (LMA-INA). The authors acknowledge Isabel Rivas for TEM sample preparation and Alfonso Ibarra for the assistance in the TEM measurements.